\documentclass[reprint,superscriptaddress,amsmath,amssymb,aps,prl]{revtex4-2}
\usepackage{graphicx}
\usepackage{siunitx}
\usepackage{ulem}
\usepackage{xcolor}
\usepackage{subcaption}
\usepackage{supertabular}
\usepackage[colorlinks=true]{hyperref}
\usepackage[utf8x]{inputenc}
\usepackage{booktabs}
\usepackage[justification=raggedright]{caption}

\begin{document}

\title{Precise Measurement of $^{216}$Po Half-life with Exact Parent-daughter Pairing in PandaX-4T}

\def\tdli{State Key Laboratory of Dark Matter Physics, Key Laboratory for Particle Astrophysics and Cosmology (MoE), Shanghai Key Laboratory for Particle Physics and Cosmology, Tsung-Dao Lee Institute \& School of Physics and Astronomy, Shanghai Jiao Tong University, Shanghai 201210, China}
\def\sjtuphys{State Key Laboratory of Dark Matter Physics, Key Laboratory for Particle Astrophysics and Cosmology (MoE), Shanghai Key Laboratory for Particle Physics and Cosmology, School of Physics and Astronomy, Shanghai Jiao Tong University, Shanghai 200240, China}
\def\newcorner{New Cornerstone Science Laboratory, Tsung-Dao Lee Institute, Shanghai Jiao Tong University, Shanghai 201210, China}
\def\MESJTU{School of Mechanical Engineering, Shanghai Jiao Tong University, Shanghai 200240, China}
\def\SPEIT{SJTU Paris Elite Institute of Technology, Shanghai Jiao Tong University, Shanghai 200240, China}
\def\SJTUSC{Shanghai Jiao Tong University Sichuan Research Institute, Chengdu 610213, China}

\def\BUAA{School of Physics, Beihang University, Beijing 102206, China}
\def\BUAACenter{Peng Huanwu Collaborative Center for Research and Education, Beihang University, Beijing 100191, China}
\def\BUAALab{International Research Center for Nuclei and Particles in the Cosmos \& Beijing Key Laboratory of Advanced Nuclear Materials and Physics, Beihang University, Beijing 100191, China}
\def\SCNT{Southern Center for Nuclear-Science Theory (SCNT), Institute of Modern Physics, Chinese Academy of Sciences, Huizhou 516000, China}

\def\USTClab{State Key Laboratory of Particle Detection and Electronics, University of Science and Technology of China, Hefei 230026, China}
\def\USTCdep{Department of Modern Physics, University of Science and Technology of China, Hefei 230026, China}

\def\YaLongSD{Yalong River Hydropower Development Company, Ltd., 288 Shuanglin Road, Chengdu 610051, China}
\def\scKeyLab{Jinping Deep Underground Frontier Science and Dark Matter Key Laboratory of Sichuan Province, Liangshan 615000, China}

\def\pku{School of Physics, Peking University, Beijing 100871, China}
\def\CHEPpku{Center for High Energy Physics, Peking University, Beijing 100871, China}

\def\SDUdep{Research Center for Particle Science and Technology, Institute of Frontier and Interdisciplinary Science, Shandong University, Qingdao 266237, China}
\def\SDUlab{Key Laboratory of Particle Physics and Particle Irradiation of Ministry of Education, Shandong University, Qingdao 266237, China}

\def\UMD{Department of Physics, University of Maryland, College Park, Maryland 20742, USA}

\def\SYU{School of Physics, Sun Yat-Sen University, Guangzhou 510275, China}
\def\SYUSFI{Sino-French Institute of Nuclear Engineering and Technology, Sun Yat-Sen University, Zhuhai 519082, China}
\def\SYUzhuhai{School of Physics and Astronomy, Sun Yat-Sen University, Zhuhai 519082, China}
\def\SYUshenzhen{School of Science, Sun Yat-Sen University, Shenzhen 518107, China}

\def\NKU{School of Physics, Nankai University, Tianjin 300071, China}
\def\YTU{Department of Physics, Yantai University, Yantai 264005, China}
\def\FDU{Key Laboratory of Nuclear Physics and Ion-beam Application (MOE), Institute of Modern Physics, Fudan University, Shanghai 200433, China}
\def\CDUT{College of Nuclear Technology and Automation Engineering, Chengdu University of Technology, Chengdu 610059, China}

\affiliation{\tdli}
\author{Chenxiang Li}\affiliation{\sjtuphys}
\author{Zihao Bo}\affiliation{\sjtuphys}
\author{Wei Chen}\affiliation{\sjtuphys}
\author{Xun Chen}\affiliation{\tdli}\affiliation{\SJTUSC}\affiliation{\scKeyLab}
\author{Yunhua Chen}\affiliation{\YaLongSD}\affiliation{\scKeyLab}
\author{Chen Cheng}\affiliation{\BUAA}
\author{Xiangyi Cui}\affiliation{\tdli}
\author{Manna Deng}\affiliation{\SYUSFI}
\author{Yingjie Fan}\affiliation{\YTU}
\author{Deqing Fang}\affiliation{\FDU}
% ...
\collaboration{PandaX Collaboration}
\noaffiliation

\date{\today}
\begin{abstract}
We report a precise measurement of $^{216}\rm Po$ half-life using the PandaX-4T liquid xenon time projection chamber (TPC).
$^{220}\rm Rn $, emanating from a $^{228}\rm Th $ calibration source, is injected to the detector and undergoes successive $\alpha$ decays, first to $^{216}\rm Po$ and then to $^{212}\rm Pb$.
PandaX-4T detector measures the 5-dimensional (5D) information of each decay, including time, energy, and 3-dimensional positions.
Therefore, we can identify the $^{220}\rm Rn $ and $^{216}\rm Po$ decay events and pair them exactly to extract the lifetime of each $^{216}\rm Po$.
With a large data set and high-precision $^{220}\rm $Rn-$^{216}\rm $Po pairing technique, we measure the $^{216}\rm Po$ half-life to be $143.7\pm0.5$~ms, which is the most precise result to date and agrees with previously published values.
The leading precision of this measurement demonstrates the power of 5D calorimeter and the potential of exact parent-daughter pairing in the xenon TPC. 
\end{abstract}

\maketitle

Since its discovery, the study of $\alpha$ decay has been essential in nuclear physics~\cite{Gamow:1928zz}.
Half-lives of the $\alpha$ decays are highly sensitive to nuclear shell effects and provide valuable insights into the prediction of the shell model, particularly for exotic nuclei (e.g., superheavy or neutron-rich/proton-rich nuclei)~\cite{Hofmann:2000cs}. 
The half-life measurement also probes the collective nuclear dynamics, such as pairing correlations and clustering phenomena~\cite{Varga:1992zz}.
In nuclear astrophysics, $\alpha$-decay rates influence the abundance patterns of heavy elements in rapid neutron-capture processes (r-process)~\cite{Kajino:2019abv}. 
Improved half-life measurements reduce uncertainties in nucleosynthesis models for events like neutron star mergers.

$^{216}\rm Po$ is a short-lived $\alpha$-emitter in the $^{228}\rm Th$ decay chain.
In the past, several measurements of the $^{216}\rm Po$ half-life have been conducted~\cite{1stPoResult,2ndPoResult,3rdPoResult,4thPoResult,5thPoResult,2017Po,2021Po}, with dedicated setup as well as with detectors specializing in double $\beta$ decay searches~\cite{5thPoResult, 2021Po}.  
The previously most precise half-life value of $144.0\pm0.6$~ms was obtained from a dedicated experiment with two silicon detectors by measuring the $\alpha$ energies~\cite{2017Po}.
Accurate pairing of $^{216}\rm Po$ with the mother nucleus $^{220}\rm Rn$ was challenging due to the high counting rate in the study and limited the measurement precision. 

Our measurement is based on the PandaX-4T liquid xenon time projection chamber (LXe TPC), which measures precise timing, energy, and 3-dimensional position of each event~\cite{PandaX-4T:2021bab}.
We name it a 5D calorimeter to highlight the unique capability. 
With all the information, Rn-Po pairs can be identified simultaneously at different detector locations. 
We perform a measurement on the decay time of $^{216}\rm Po$ with significant statistics during the calibrations for low energy electronic recoil (ER) events with internal radon sources~\cite{Ma:2020Rn}.  
Our half-life measurement represents one of the first competitive nuclear physics results produced by a LXe TPC and demonstrates the power of the 5D calorimeter.
\begin{figure}[tb]
    \centering
    \includegraphics[width=\columnwidth]{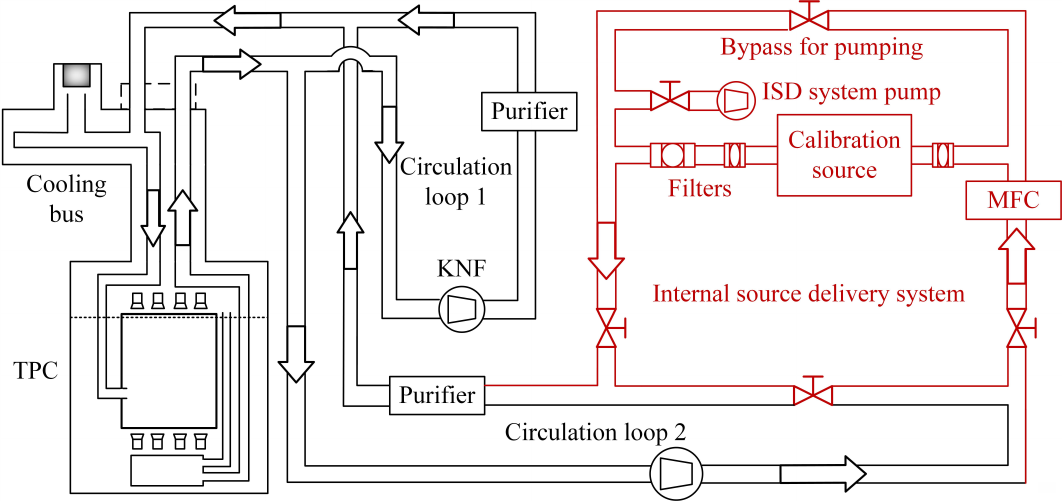}
    \caption{PandaX-4T TPC and xenon circulation system (in black) with the internal source delivery system (in red).}
    \label{fig:PandaX}
\end{figure}

The PandaX-4T experiment is located at the China Jin-Ping Underground Laboratory (CJPL) in Sichuan Province, China~\cite{Jinping, CJPL2, PhysRevD.93.122009}.
The experiment utilizes a series of xenon TPCs for dark matter direct detection and neutrino physics~\cite{PandaXB8Neutrino, PandaX:2024qfu,PandaX:2023ggs}. 
A schematic drawing of key PandaX components is shown in Fig.~\ref{fig:PandaX}, with the internal calibration source delivery system highlighted. 
The sensitive volume (SV) of the TPC contains 3.7 tons of liquid xenon. 
Any sizable energy deposited in the SV, for example, $\alpha$ particles emitted from $^{216}\rm Po$ decays, would cause xenon to scintillate and ionize.
Ionized electrons drift upward in a downward-pointing electric field in the field cage.
When reaching the liquid-gas interface, the electrons are extracted into the xenon gas and produce photons via the electroluminescence under a more intensive electric field.
The prompt scintillation signals (usually called $S1$) and delayed electroluminescence signals ($S2$) are captured by the top and bottom photomultiplier tube (PMT) arrays.
The timing of each event is determined from the timestamp of $S1$.
The energy can be calculated from the number of photoelectrons (PEs) collected by PMT arrays for $S1$ and $S2$, denoted as $qS1$ and $qS2$, respectively~\cite{Yang_2022}.

$S2$ photons are emitted a few centimeters from the top PMT array, and the light collection pattern on top PMTs gives the horizontal \textit{X-Y} position of the event.
We have developed multiple algorithms for position reconstruction, and the relative precision is in the order of a few millimeters in \textit{X-Y} and sub-millimeter in \textit{Z} direction~\cite{PAF}.
The vertical \textit{Z} position can be precisely determined by multiplying the time delay between $S1$ and $S2$ by drift velocity, which is 1.4~mm/$\mu$s at the drift field of 13.5~V/mm when we took the radon calibration data.

$^{216}\rm Po$ is introduced to PandaX-4T detector as an intermediary nucleus of $^{228}\rm Th $ decay chain $ {^{228}\rm Th} \rightarrow {^{224}\rm Ra} \rightarrow {^{220}\rm Rn} \rightarrow {^{216}\rm Po} \rightarrow {^{212}\rm Pb} \rightarrow {^{212}\rm Bi}$.
$^{220}\rm Rn $ and its daughters in the decay chain have a relatively short lifetime, making it suitable for internal calibration in search for dark matter and other rare events with LXe TPC.
$\beta$ decays of $^{212}\rm Pb$ are commonly used to characterize the detector response to electronic recoil signals.
A thousand-Bq level $^{228}\rm Th $ source electroplated on a stainless steel disk is utilized in PandaX.
$^{220}\rm Rn$ in the chain may emanate from the disk surface and get delivered to the detector via an internal source delivery system (ISDS), as shown in Fig.~\ref{fig:PandaX}.

The ISDS is appended to one of the xenon gas circulation loops, as shown in Fig.~\ref{fig:PandaX}.
The xenon circulation loops extract xenon from the TPC, remove electronegative impurities with high-temperature purifiers, and liquefy xenon again before sending it back to the detector~\cite{Zhao_2021,distillation,Cui_2024}. 
During the calibration, the ISDS is connected to the circulation loop, and xenon gas flows through the ISDS to carry the emanated $^{220}\rm Rn$ to the detector.
The main components of the ISDS include a source chamber, particulate filters, and a mass flow controller (MFC). 
The $^{228}\rm Th $ source disk is mounted in a stainless steel cylindrical chamber with a diameter of 10~cm and length of 20~cm.
Two particulate filters rated at 2~$\mu$m and 3~nm are installed downstream of the source chamber, preventing solid radioactive dust (e.g. $^{228}\rm Th $) larger than the rated values from entering the TPC. 
A 2-$\mu$m filter is also installed upstream to avoid possible back-flow of the radioactive particulates.
The MFC controls the xenon gas circulation rate through the source chamber and, therefore, the amount of $^{220}\rm Rn$ entering the TPC.
During a calibration campaign in May 2021, a $^{220}\rm Rn$ activity of approximately 1.7~Bq was introduced in the TPC.
No noticeable long-lived radioactive impurities were introduced during the calibration.
In this analysis, we used a total of 35.4 hours of data during this calibration campaign.

\begin{figure}[tb]
    \centering
    \includegraphics[width=1\columnwidth]{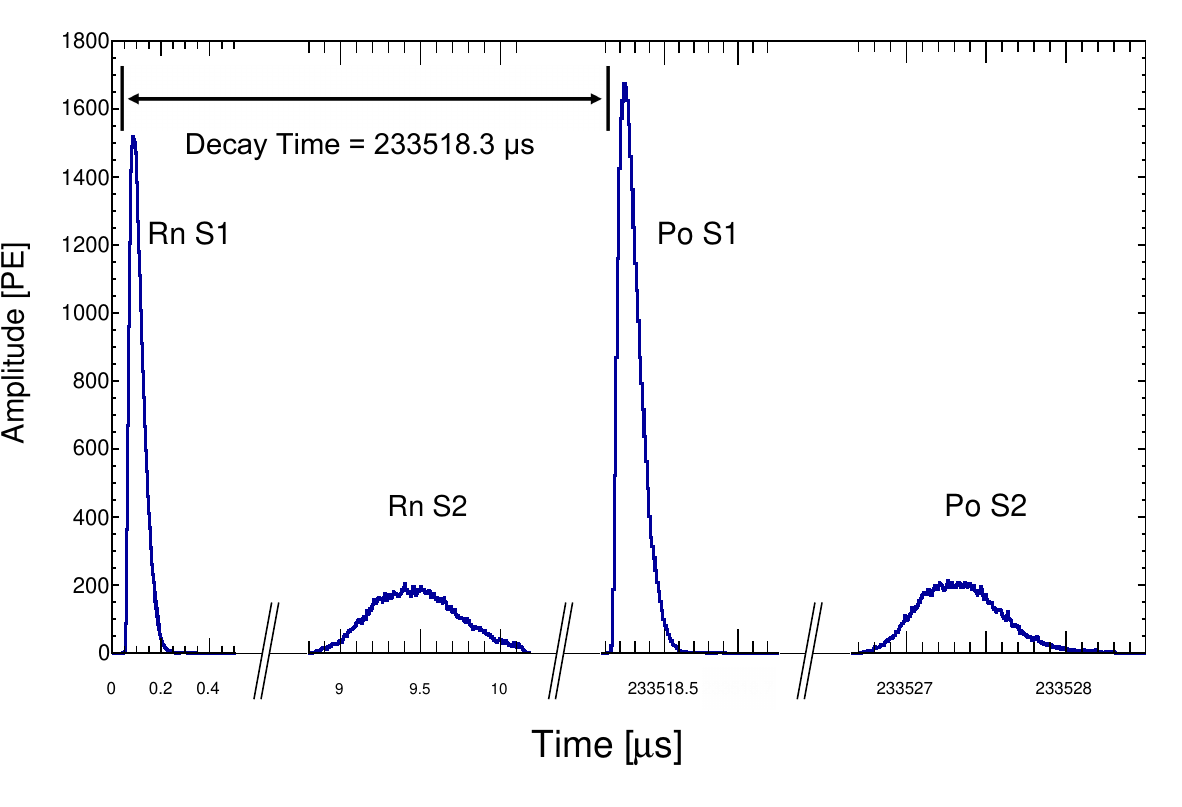}
    \caption{An example time series of $^{220}\rm Rn$ and $^{216}\rm Po$ pair.
 The four pulses are labeled as in the figure. While the scale for signals of the same type remains consistent, the S2 scale is larger than the S1 scale.
 The lifetime of the Po nucleus for this example is approximately 233~ms.}
    \label{fig:signal}
\end{figure}

A typical $^{220}\rm Rn$-$^{216}\rm Po$ (will be abbreviated as Rn-Po later) pair with two $\alpha$ events in the SV is shown in Fig.~\ref{fig:signal}.
The first peak is the $S1$ signal from the $^{220}\rm Rn$-emitted $\alpha$ particle, followed by its $S2$ of the same event.
In the figure, the $S1$ signal from the daughter $^{216}\rm Po$ is approximately 233~ms after the Rn $S1$ signal, which gives the lifetime of this particular $^{216}\rm Po$.

\begin{figure}
    \centering
    \includegraphics[width=\columnwidth]{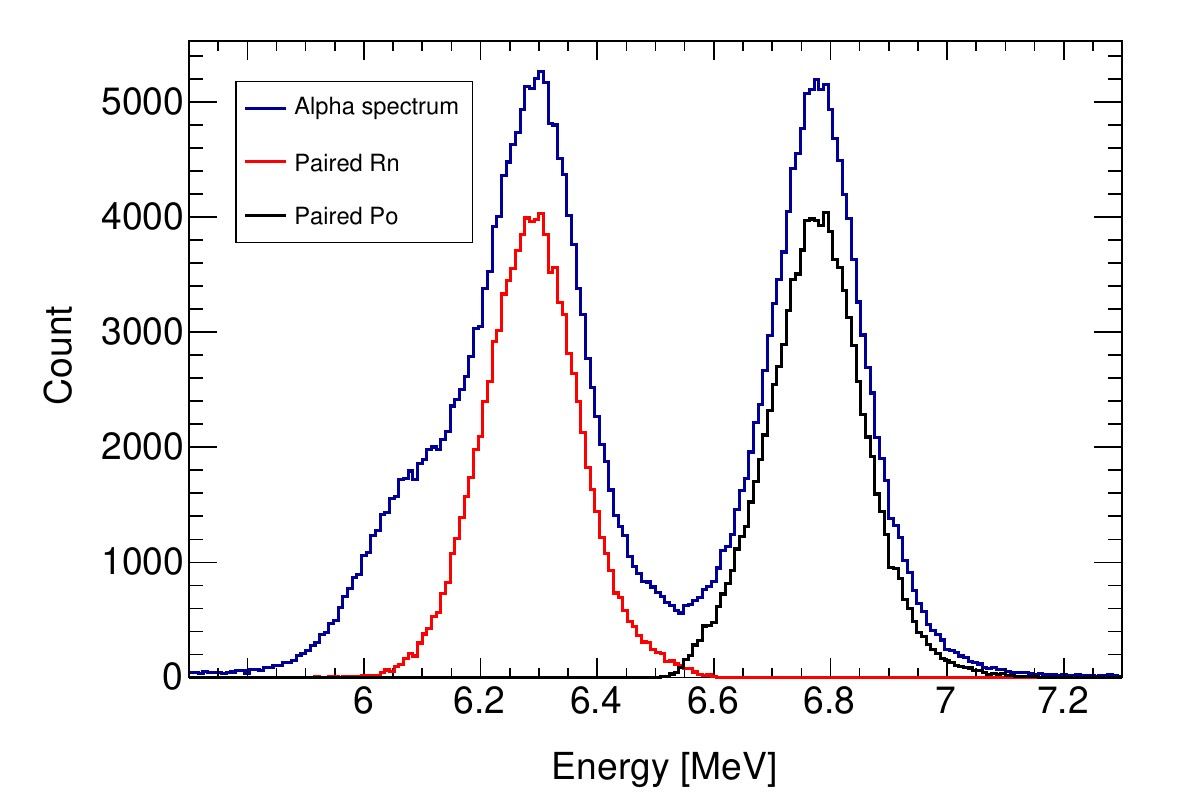}
    \caption{Energy spectra of $\alpha $ events in SV before (in blue) and after pairing (in red and black). 
 Two highest peaks are from $\alpha$ decays of $^{220}\rm Rn $ and $^{216}\rm Po $, as marked in the figure. $^{212}\rm Bi $ from the decay chain after $^{216}\rm Po $ are removed efficiently by the Rn-Po pairing. }
    \label{fig:Spectrum}
\end{figure}

The $\alpha$ decay events from the Rn-Po pair can be selected with energy cuts.
The energies for $^{220}\rm Rn $ and $^{216}\rm Po $ $\alpha$s are 6.288~MeV and 6.778~MeV, respectively, which are higher than that of most of the naturally occurring $\beta$ and $\gamma$ particles.
Therefore, $\alpha$ particles can be relatively straightforwardly identified with energy selection criteria.
High-energy $\alpha$ particles deposit energy in xenon with much larger $dE/dx$ than $\beta/\gamma$ particles. 
The recombination of ionization electrons with local positive ions before drifting results in smaller $qS2$ and larger $qS1$ compared to $\beta/\gamma$ events in the detector.
The pulses in Fig.~\ref{fig:signal} demonstrate the relative sizes of $qS1$ and $qS2$, which are used more efficiently to select $\alpha$ particles.
Also, because of the relatively large scintillation light yield, $qS1$ is used to reconstruct the energies of $\alpha$ events.
The blue curve in Fig.~\ref{fig:Spectrum} shows the spectrum of high energy $\alpha$s.
The two main dominant peaks are from $^{220}\rm Rn $ and $^{216}\rm Po $ with a minor peak from $^{212}\rm Bi $ on the left shoulder of the $^{220}\rm Rn $ $\alpha$ peak.
The relative number of $\alpha$ events between $^{216}\rm Po $ and $^{212}\rm Bi$ is a reflection of $^{212}\rm Bi$ decay branches and the depletion effect of radon decay chain in liquid xenon TPC~\cite{Ma:2020Rn, Nedlik:2022fdx, Jorg:2023nvl}.

The lifetime of each $^{216}\rm Po $ nucleus can be accurately determined from the timing information of $\alpha$ decays.
The timestamp of $^{220}\rm Rn $ ($^{216}\rm Po $) $\alpha$ event corresponds to the formation (decay) of a $^{216}\rm Po $ nucleus. 
In the LXe TPC, the timestamp of each event is determined from $S1$ scintillation signals.
The $S1$ from an $\alpha$ event has a typical width, which covers 90$\%$ of the whole signal charge, of around 0.2~$\mu$s and a peak timing uncertainty of a few nanoseconds, which are much smaller than the expected half-life and can be neglected.
Once we pair the consequent Rn-Po $\alpha$ decays, the time difference between two $S1$ signals indicates the lifetime of the $^{216}\rm Po $.


An accurate measurement of $\alpha$ positions in the LXe TPC facilitates the exact pairing of Rn-Po, even under the high decay event rate.
Particles may travel in the TPC due to xenon convection, but the velocity is measured on a scale of millimeters per second~\cite{Malling:2014oxk,XENON:2016RnField,XENONCollaborationP:2024xwn}.  
Considering the relatively short half-life of $^{216}\rm Po $, the distance between Rn and Po $\alpha$ events can be taken as a restrictive condition for Rn-Po pairing.

With the energy, timing, and position of $^{220}\rm Rn $ and $^{216}\rm Po $, the exact pairing of the parent and daughter nuclei can be performed as follows.
Firstly, $\alpha$ events in the SV of the detector are identified as $^{220}\rm Rn $ or $^{216}\rm Po $ based on energy, as shown in Fig.~\ref{fig:Spectrum}. 
Energy reconstruction has fluctuations in the \textit{Z} direction. Instead of separating $^{220}\rm Rn $ or $^{216}\rm Po $ events simply by the valley value of two main peaks in the overall spectrum in Fig.~\ref{fig:Spectrum}, the split is done in multiple spectra at different \textit{Z} positions to mitigate the bias of energy reconstruction on the \textit{Z} position.

Secondly, to avoid bias when estimating the $^{216}\rm Po $ half-life, we select $^{220}\rm Rn $ events only from the fiducial volume (FV) within the SV, while including $^{216}\rm Po $ pairing events from the entire SV. 
The determination of the FV boundary primarily arises from two considerations: the $^{216}\rm Po $ leak-out and data quality concerns within certain region.
For a $^{220}\rm Rn $ event that occurs near the SV boundary, its daughter $^{216}\rm Po $ may travel outside of the detector, especially if the lifetime of this specific nucleus is long. 
Therefore, the leak-out may introduce a bias towards a shorter half-life.
So the first cut of FV is 10 cm inward from the side and 1 cm from the top and bottom of the SV's boundary. 
The second cut of FV removes the events from the upper right corner of in the detector, which is called the top-dense region.
Most radon nuclei enter the TPC via this region, and many decay here as well. 
The distribution of events in the top-dense region reveals an unphysical clustering underneath PMT active areas.
Therefore, we remove any events in the top 70~mm of the liquid xenon in the top-dense region to avoid mis-pairing due to problematic position reconstructions.
The FV cut on $^{220}\rm Rn $ events is the main reason the number of paired Rn-Po events is smaller than the original spectrum.

Thirdly, a maximum time difference of 3 seconds is used as a restrictive condition for Rn-Po pairing. 
Since the time cut is more than 20 times the expected $^{216}\rm Po $ half-life, the possibility of missing a true Rn-Po is negligible.
Considering the average $\alpha$ rate of approximately 3.4~Hz in the FV, additional position information is important for the pairing.

Finally, the spatial proximity of Rn and Po events is used to pair them exactly.
A candidate $^{216}\rm Po $ daughter must occur within 1.5~cm vertically and 5~cm horizontally from the $^{220}\rm Rn $ parent.
The vertical cut is determined based on the commonly measured xenon convection velocity of around 3~mm/s for meter-scale detector~\cite{XENONCollaborationP:2024xwn}.
The horizontal cut is looser to accommodate the degraded position resolution in the direction.
We also evaluate $^{216}\rm Po$ half-lives based on different proximity cuts and treat the difference in results as systematic uncertainties.

With all the cuts applied, around 3.4\% of $^{220}\rm Rn $ candidates event end up with multiple $^{216}\rm Po$ matches. 
Exact pairing is impossible for these $^{220}\rm Rn$ candidates, regardless of whether we choose $^{216}\rm Po$ counterparts based on the closest distance or minimum time interval. 
Consequently, they are removed from the final half-life fits.

\begin{figure}
    \centering
    \includegraphics[width=\columnwidth]{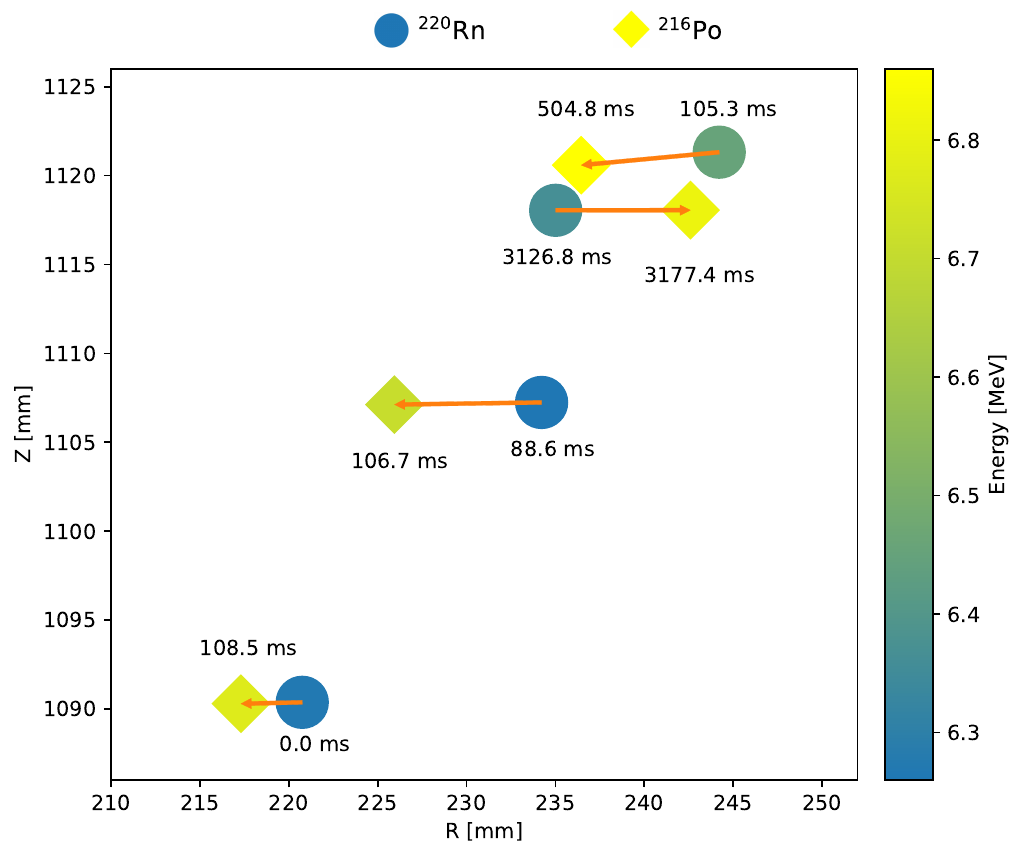}
    \caption{Example of exact Rn-Po pairing. 
 Blue circles represent $^{220}\rm Rn$, green circles represent $^{216}\rm Po$, and the circle radii indicate energy differences.
 Yellow arrows link the Rn-Po pairs. 
 The value marked near each circle (in ms) indicates the time interval between it and its parent Rn event.}
    \label{fig:Pairing}
\end{figure}

The energy spectra of paired Rn and Po events are shown in red and black in Fig.~\ref{fig:Spectrum}.
As an illustration of the efficiency of our 5D cuts, a total of 98948 pairs are selected out of 147302 $^{216}\rm Po$ $\alpha$ events.
Four pairs of Rn and Po in a selected volume are shown in Fig.~\ref{fig:Pairing}.
In the figure, the relative $\alpha$ decay timestamps in milliseconds are marked by the event and the sizes of the circles illustrate the energy of the $\alpha$ event. 
The temporally overlapped events are well separated in space and vice versa.
The figure illustrates the usefulness of 5D calorimetry in Rn-Po pairing.

\begin{figure}
    \centering
    \includegraphics[width=\columnwidth]{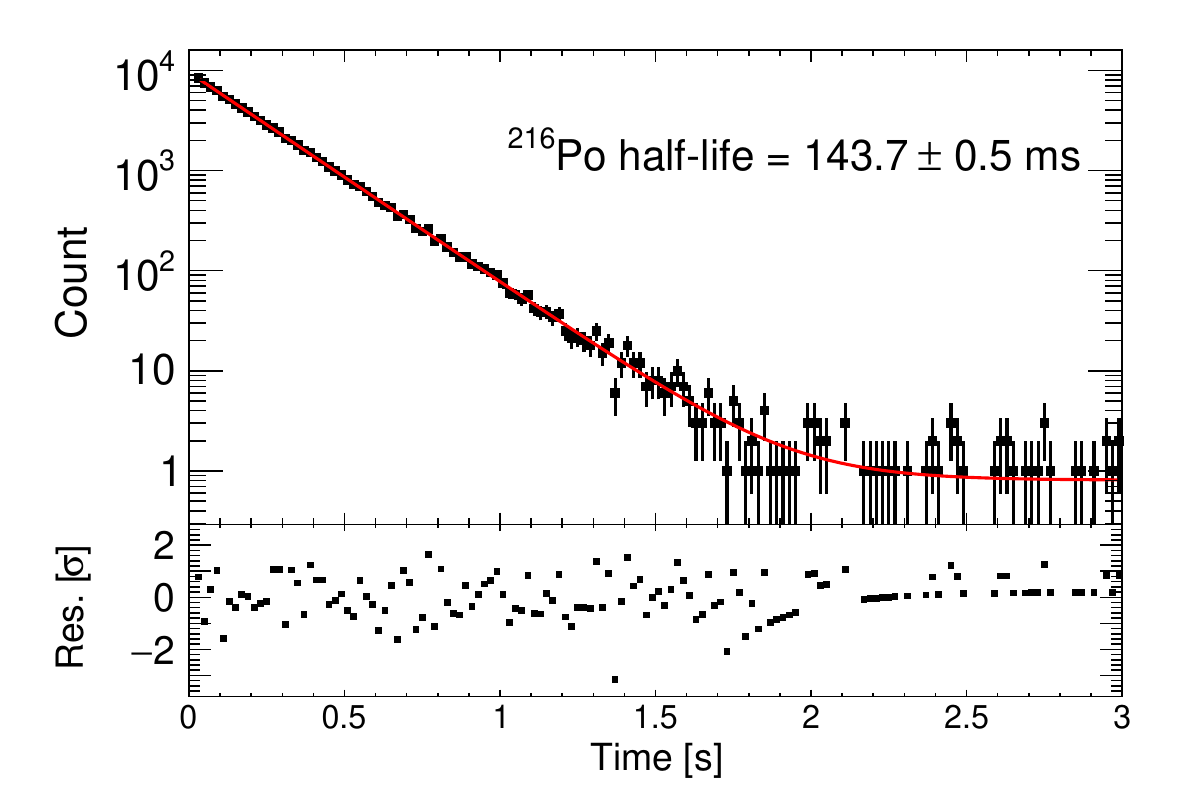}
    \caption{The distribution of $^{216}\rm Po $ decay time. 
 The red fit curve is an exponential plus a constant function with three free parameters in total.
 The lower panel shows residuals in the unit of $\sigma$, defined as residuals divided by the square root of the number of events in the bins.}
    \label{fig:FitResults}
\end{figure}

The half-life value of $^{216}\rm Po $ is obtained from a total of 98948 Rn-Pn pairs after all the cuts.
The distribution of $^{216}\rm Po $ lifetimes and corresponding fit are shown in Fig.~\ref{fig:FitResults}.
The histogram is fitted with the sum of an exponential curve with the decay half-life and a flat background.
The background is introduced to account for falsely paired random events by coincidence, which have even-distributed lifetimes. 
A binned maximum likelihood method is used for the fit, resulting in a half-life of 143.7~ms with a statistical uncertainty of 0.5~ms.

\begin{figure}
    \centering
    \includegraphics[width=\columnwidth]{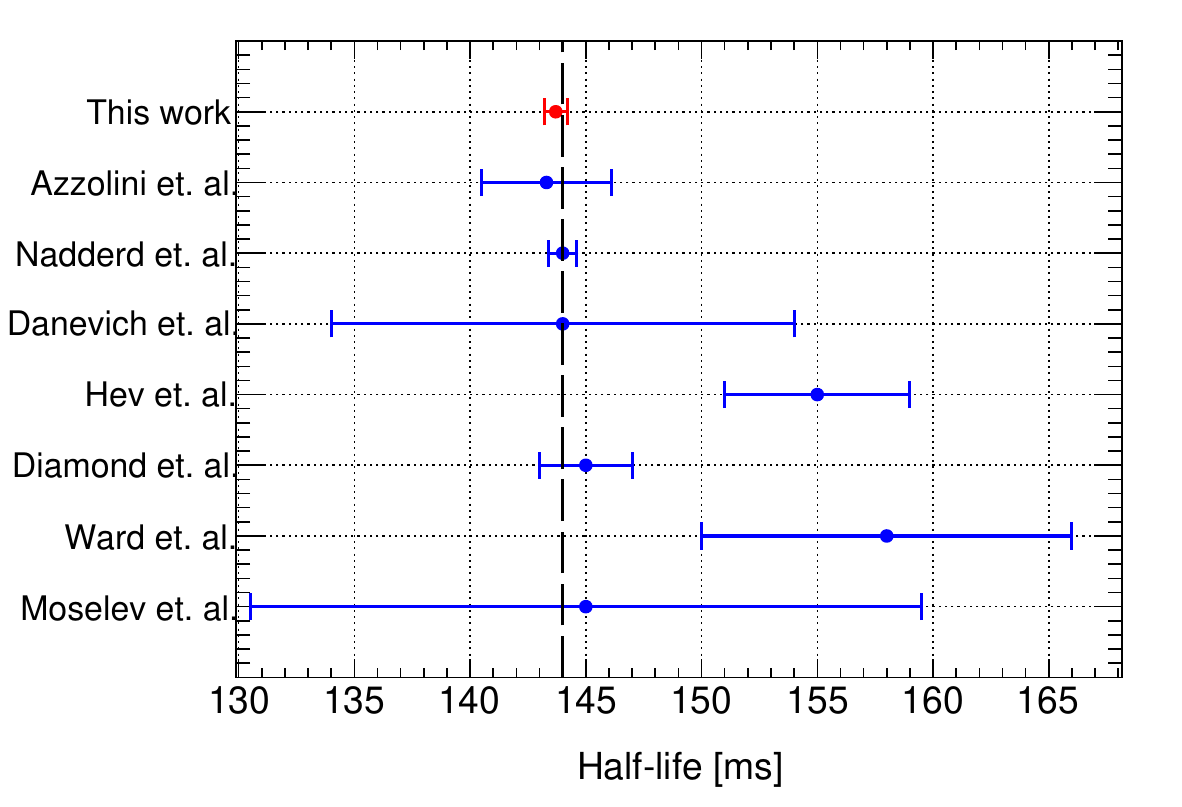}
    \caption{Our half-life value (in red) is compared with previous results (in blue)~\cite{1stPoResult,2ndPoResult,3rdPoResult,4thPoResult,5thPoResult,2017Po,2021Po}.
 The dashed line represents the combined half-life when our result is included.}
    \label{fig:Previous}
\end{figure}

\begin{table}[!hpt]
  \caption{Systematic uncertainties components.}
  \label{tab:Err}
  \centering
  \begin{tabular}{cc} \toprule
    Component & Uncertainties (ms)  \\ \midrule
    Data selection    & 0.06 \\
    Fiducial volume      & 0.13 \\
    Pairing      & 0.09 \\
    Fit procedures  & 0.12 \\ \bottomrule
  \end{tabular}
\end{table}

Systematic uncertainties from the data selection, fiducial volume, pairing, and fit procedures in Table.~\ref{tab:Err} are considered and discussed in detail below.
We iterate different scenarios and compare the fit results with our baseline fit value.
A total systematic uncertainty of 0.2~ms is counted. 

Misidentification of Rn and/or Po $\alpha$ events introduces systematic uncertainties.
The presence of $^{212}\rm Bi$ in $^{220}\rm Rn$ can not be avoided due to limited energy resolution, as shown in Fig.~\ref{fig:Spectrum}.
Additionally, it is possible to have Rn or Po $\alpha$ particles reconstructed at the wrong energy peak and mis-tagged. 
$^{220}\rm Rn$ events with energy within the $E_{0}\pm 1\sigma$ to within $E_{0}\pm 3\sigma$ are selected for different pairing scenarios, where $E_{0}$ and $\sigma$ are the mean and width values calculated from a Gaussian fit of the Rn peak.
Half-life values from traversed $^{220}\rm Rn$ selections differ from our baseline fit result by at most 0.06 ms, which is treated as a systematic uncertainty.

The uncertainty introduced by the choice of FV boundaries is 0.13 ms.
For the first cut of FV, we iteratively tested boundary vertical intervals between FV and SV from 1 cm to 5 cm in 1 cm increments, which introduced an uncertainty of 0.09 ms.
Similarly, we tested horizontal interval squares between FV and SV from 2500 cm² to 2000 cm² in 100 cm² decrements, introducing an uncertainty of 0.04 ms.
For the second cut of FV, we iterate the bottom boundary of the top-dense region from 70 to 140 mm below the liquid level in steps of 14 mm, and the uncertainty introduced is 0.03 ms.
Similarly, we iterate the horizontal boundary in steps of 14~mm and obtain a maximum difference of 0.08~ms in half-life values.

The uncertainty introduced by spatial proximity cuts in pairing on the pairing procedure is 0.09~ms. 
We iterate the horizontal distance cut from 35 to 70~mm with a step of 5~mm.
The maximum difference between the fit results of the iteration and our baseline value is 0.05~ms.
Similarly, the vertical direction cut is iterated from 10 to 22 mm by every 2 mm, and an uncertainty of 0.07ms is introduced. 

Uncertainties introduced from the alternative fit parameters amount to 0.12~ms. 
We iterate the bin width for 4 to 20~ms, and the fit half-lives are at most 0.04~ms different from the baseline result. 
The most significant difference in fit half-life by varying the fit start point from 10~ms to 70~ms is 0.11~ms. 
Both values are treated as systematic uncertainties from the fit. 

After considering the statistical uncertainties and add all systematic uncertainties in quadrature, our final measured result of $^{216}\rm Po $ half-life is $143.7\pm0.5 \,\rm{(stat.)}\pm0.2\, \rm{(syst.)}$~ms, dominated by the statistical one. 
Our result is the most precise measurement of $^{216}\rm Po $ half-life to date and agrees with the previous three precise measurements within one sigma~\cite{2017Po,2021Po,3rdPoResult}, as shown in Fig.~\ref{fig:Previous}.
Comparing with two other $^{216}\rm Po $ half-lives from underground rare event experiments~\cite{5thPoResult, 2021Po}, we have improved uncertainty by a factor of 20 and 5.6 respectively. 
The average half-life of $^{216}\rm Po $ only from previous results is $144.3\pm0.6$~ms.
If we combine our value and previous results (Fig.~\ref{fig:Previous}), the new average half-life of $^{216}\rm Po $ is $144.0\pm0.4$~ms. 
The uncertainty has also been improved.

In summary, a most precise measurement of $^{216}\rm Po $ half-life has been achieved in this work, thanks to the accurate measurement of time, energy, and 3D position in the PandaX-4T TPC.
We have determined the starting and ending of all $^{216}\rm Po $ nuclei unambiguously by exact Rn-Po pairing. 
The precise measurement demonstrates the unique capability of liquid xenon TPC as a 5D calorimeter.
The exact pairing techniques have a wide variety of applications in the convection field studies~\cite{XENONCollaborationP:2024xwn,EXO-200:2015Rn}, position reconstruction validations~\cite{XENON:2016RnField}, and background modeling~\cite{XENONCollaborationP:2024xwn,Jorg:2022iwn}. 
For convection field studies, $^{222}\rm Rn $-$^{218}\rm Po $ pairing can reconstruct the LXe flow in the entire meter-scale detector because the half-life of $^{218}\rm Po $ is sufficiently long.
For the validation of position construction, each $^{220}\rm Rn $–$^{216}\rm Po $ pair can provide differential position values in vertical and horizontal directions. 
For background modeling, after obtaining the complete LXe convection field from $^{222}\rm Rn $–$^{218}\rm Po $ pairing, $^{214}\rm Pb $ background events can be tagged by the following $^{214}\rm Bi $ and $^{214}\rm Po $ decays, or by the preceding $^{218}\rm Po $ decays.

Further developments in half-life measurements in LXe TPCs are also anticipated. 
A promising candidate is pairing $^{212}\rm Bi $ and $^{208}\rm Tl $ to precisely measure the $^{208}\rm Tl $ half-life of about 3 minutes using $^{220}\rm Rn $ calibration data. 
Dedicated campaigns to introduce candidate isotopes into the TPC would further expand its applications in nuclear physics.

\section*{Acknowledgment}
We thank Yi-Fei Niu for helpful discussions on the topic. 
This project is supported in part by grants from National Key R$\&$D Program of China (Nos. 2023YFA1606200, 2023YFA1606202), National Science Foundation of China (Nos. 12090060, 12090062, U23B2070), and by Office of Science and Technology, Shanghai Municipal Government (grant Nos. 21TQ1400218, 22JC1410100, 23JC1410200, ZJ2023-ZD-003). We thank for the support by the Fundamental Research Funds for the Central Universities. We also thank the sponsorship from the Chinese Academy of Sciences Center for Excellence in Particle Physics (CCEPP), Hongwen Foundation in Hong Kong, New Cornerstone Science Foundation, Tencent Foundation in China, and Yangyang Development Fund. Finally, we thank the CJPL administration and the Yalong River Hydropower Development Company Ltd. for indispensable logistical support and other help. 

\normalem
\bibliography{article}
\end{document}